\documentclass[manuscript]{acmart}
\AtBeginDocument{%
  \providecommand\BibTeX{{%
    \normalfont B\kern-0.5em{\scshape i\kern-0.25em b}\kern-0.8em\TeX}}}

\usepackage{fontawesome5}

\setcopyright{acmlicensed}
\copyrightyear{2024}
\acmYear{2024}

\acmConference[CHI'24]{TREW workshop (CHI'24)}{2024}{Hawaii, USA}
\usepackage[]{mdframed}

\usepackage{enumitem}

\newcommand{\nb}[3]{
  \fcolorbox{black}{#2}{\bfseries\sffamily\scriptsize#1}
    {\sf\small$\blacktriangleright$\textit{\textcolor{brown}{#3}}$\blacktriangleleft$}
}

\definecolor{lavendermist}{rgb}{0.9, 0.9, 0.98}
\definecolor{aliceblue}{rgb}{0.94, 0.97, 1.0}
\definecolor{antiquewhite}{rgb}{0.94, 1, 1}

\newcommand\agathe[1]{\nb{Agathe}{green}{#1}}

\newcommand\mireia[1]{\nb{Mireia}{yellow}{#1}}

\begin{document}

\title[An Empirical Exploration of Trust Dynamics in LLM Supply Chains]{
An Empirical Exploration of Trust Dynamics in LLM Supply Chains
}

\author{Agathe Balayn}
\affiliation{%
  \institution{ServiceNow, Trento University, Delft University of Technology}
  \country{The Netherlands}
}
\email{a.m.a.balayn@tudelft.nl}

\author{Mireia Yurrita}
\email{m.yurritasemperena@tudelft.nl}
\affiliation{%
  \institution{Delft University of Technology}
  \country{The Netherlands}
}

\author{Fanny Rancourt}
\email{fanny.rancourt@servicenow.com}
\affiliation{%
  \institution{ServiceNow}
  \country{Canada}
}

\author{Fabio Casati}
\affiliation{%
  \institution{ServiceNow}
  \country{Switzerland}}
\email{fabio.casati@servicenow.com}

\author{Ujwal Gadiraju}
\affiliation{%
  \institution{Delft University of Technology}
  \country{The Netherlands}
}
\email{u.k.gadiraju@tudelft.nl}

\renewcommand{\shortauthors}{Balayn, et al.}

\begin{abstract}
With the widespread proliferation of AI systems, \textit{trust in AI} is an important and timely topic to navigate. Researchers so far have largely employed a myopic view of this relationship. In particular, a limited number of relevant trustors (e.g., end-users) and trustees (i.e., AI systems) have been considered, and empirical explorations have remained in laboratory settings, potentially overlooking factors that impact human-AI relationships in the real world. In this paper, we argue for broadening the scope of studies addressing `\textit{trust in AI}' by accounting for the complex and dynamic supply chains that AI systems result from. AI supply chains entail various technical artifacts that diverse individuals, organizations, and stakeholders interact with, in a variety of ways.
We present insights from an in-situ, empirical study of LLM supply chains. Our work reveals additional types of trustors and trustees and new factors impacting their trust relationships. These relationships were found to be central to the development and adoption of LLMs, but they can also be the terrain for uncalibrated trust and reliance on untrustworthy LLMs. Based on these findings, we discuss the implications for research on `trust in AI'. We highlight new research opportunities and challenges concerning the appropriate study of inter-actor relationships across the supply chain and the development of calibrated trust and meaningful reliance behaviors. We also question the meaning of building trust in the LLM supply chain. 
\end{abstract}



\maketitle
\section{Introduction}
Trust is a fundamental necessity for collaboration~\cite{vangen2003nurturing}.
For that reason, HCI researchers have started investigating trust relationships between humans and AI \cite{ueno2022trust,vereschak2021evaluate,lai2021towards}.
Such research has primarily explored the conditions for the development of appropriate trust in  AI systems \cite{baidoo2023education,tan2023generative,zheng2024judging}, for society to leverage their potential (e.g., as decision-support systems \cite{bansal2019beyond}) while avoiding their harmful impacts when inappropriately developed and used \cite{balayn2021beyond}.
To this end, it has limited the trustee (i.e., the actor that is trusted) to the AI system (that makes predictions).  In turn, the trustor, i.e., the trusting actor of such an AI system, has been limited to a human entity, whether it be the user of the AI system 
\cite{kim2023humans,vodrahalli2022humans,leichtmann2023effects}, the decision-subject who might have to decide whether to contest the decisions taken \cite{araujo2020ai} or the general public (trust in AI technologies) \cite{araujo2020ai,knowles2021sanction}.

However, the collaboration between humans and AI is not limited to the time they interact to conduct tasks collaboratively.
The development and deployment of AI systems are also a result of collaborative work between many actors in the AI supply chain \cite{cobbe2023understanding,crawford2018anatomy}. For instance, an AI system such as an LLM requires individuals in an organization to develop a foundation model, the same or another organization to fine-tune such models and deploy them in an application, a consumer organization should then be willing to adopt this application, and then the end-users can finally interact, engage, or collaborate with the LLM-powered application to carry out tasks.  
Trust dynamics along the AI supply chain can, therefore, govern or influence the nature of collaboration at each juncture. Coming back to our example, one can for instance imagine that the consumer organization first develops trust in the foundation model or fine-tuned model to decide to adopt the LLM-powered application. 


In this paper,  
we argue that trust between humans and AI should not be limited to the study of the final interaction, and should also account for the complexities of the supply chain that AI systems result from. 
Besides decades of organizational psychology research \cite{tan2009trust,lee2004trust,bach2022systematic,salam2017mediating,kwon2004factors,tejpal2013trust,pirson2009facing,bachmann2015repairing,mukherjee2012development,gwebu2007conceptual,tan2009trust}, the few in-situ studies of trust in AI have already hinted at the existence of many trust relations within AI lifecycles inside the supply chain \cite{browne2022trust} and many organizational factors impacting these relations \cite{kim2023humans}. 
It is hence evident that trust plays a key role not only in the adoption and use of AI systems but also in other synergies within the AI supply chain.

Ultimately, trust dynamics can influence the production of AI systems and their adoption, impacting the trustworthiness of AI systems employed in society, how appropriate user interactions with AI systems are, how responsible their production is, and therefore the direction in which the AI industry is headed. 
That is why we build on valuable prior research on trust in AI and and aim to broaden the scope of understanding trust relations that influence downstream collaborations and outcomes in the AI supply chain.

In this paper, we explore and identify 
\emph{the actors beyond the end user and the AI system that serve as trustors and trustees in the AI supply chain, and the factors that impact trust relationships relevant to these additional actors.}
To this end, we report preliminary findings from a study investigating trust relationships within LLM supply chains. We focus on LLM supply chains due to their complexity \cite{hacker2023regulating} and timely relevance. 
Our results illustrate the importance, diversity, and complexity of trust relations beyond known trustors and trustees, both within and across organizations. 
In the future, going beyond a myopic understanding of AI trust will require accounting for new factors impacting trust relations (e.g., organizational reputation and historical relations), and developing new metrics for understanding trust and trust calibration. 
We hope that this paper acts as a call for AI trust researchers to broaden the scope of their work to address pressing issues in unaddressed but highly relevant trust relations along AI supply chains.  


\section{Brief Overview of the Background and Methodology}

\textit{Background.}
Despite trust not having an agreed-upon definition, a majority of work agrees to characterize a relation of trust by the \textbf{positive expectation} the trustor has for the trustee, and the trustor's \textbf{vulnerability} and uncertainty vis-a-vis the trustee in the context at hand \cite{lee2004trust}.
According to prior literature, trust relations are not only characterized by the \textbf{trustor} and the \textbf{trustee}, but also by the trust \textbf{activity} \cite{smart2021risk}. For instance, the user of an LLM might trust it to (activity 1) summarize texts that do not require expertise, but might refrain from relying on it when they need to (activity 2) summarize scientific papers. 
The literature further distinguishes between trust that is or not \textbf{appropriate} and that does or not lead to a reliance and compliance behavior of the trustor on the trustee \cite{mehrotra2023systematic}.
Finally, researchers within and outside AI have already identified several \textbf{factors} that impact trust  \cite{liao2022designing,mayer1995integrative}. Existing works have explained that trust depends on a) the trustee's characteristics (particularly the ABI framework consisting of ability, benevolence and integrity \cite{mayer1995integrative}) mediated by the affordances that expose these characteristics to the trustor \cite{liao2022designing}, and on b) the trustor's characteristics (e.g., disposition for trust). c) The context of the trust activity further mediates the trust relation, e.g., impacting the trustor's systematic and heuristic processing of the trustor's affordances \cite{liao2022designing}.

In the context of AI, trust has been investigated between a user and an AI system conceptualized as one entity that makes individual predictions on input samples \cite{he2022walking}. However, AI systems, especially LLMs, are much more complex artifacts \cite{cobbe2023understanding}. We therefore intend to broaden the scope and look into trust across actors in the LLM supply chain.

\begin{figure}
    \centering
    \includegraphics[width=0.92\linewidth]{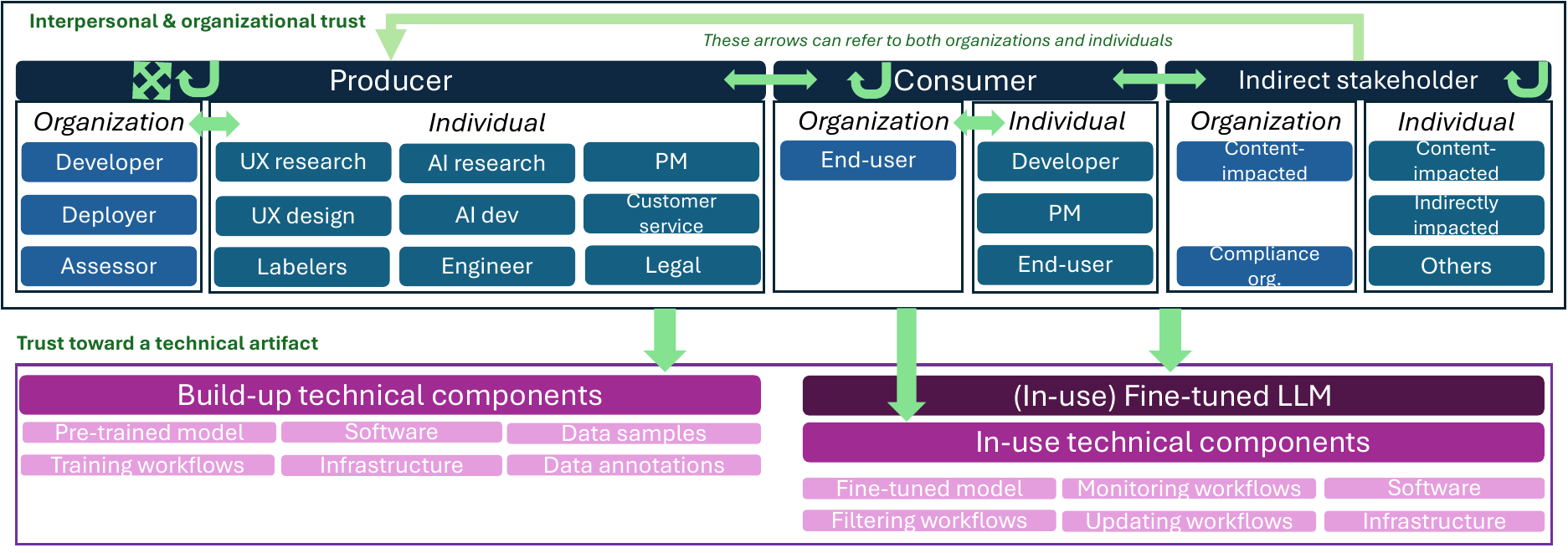}
    \caption{The primary entities (boxes) and trust relations (arrows) that shape the supply chain. 
    The 
    cross and 
    loop arrows respectively represent trust relations across different entities in the same box (e.g., developer and deployer organization), and trust relations across entities of the same type (e.g., two different developer organizations). Remember that the supply chain is not only made of trust relations, but also other relations across entities 
    (e.g., one developer organization might develop a pre-trained model and another one might adopt this model to fine-tune the LLM.)}
    \label{fig:trustSummary}
\end{figure}

\textit{Methodology.}
We conceptualize trust in LLM supply chains by combining prior work in organizational psychology, trust in AI, and the empirical insights we generated from our qualitative study. 
We are convinced of the necessity to bring more qualitative methods to the trust-in-AI research and especially more empirical \emph{in-situ} investigations of the LLM supply chain and the trust relationships that shape it. 
The few real-world, empirical, research projects about trust \cite{kim2023humans,browne2022trust} and their insights have hinted at such importance in the past.
Hence, we conducted semi-structured interviews with 71 practitioners across 10 organizations that develop, deploy, or use LLMs. Through a mixture of snowball and convenience sampling, we recruited practitioners who play different roles across the supply chain and vis-a-vis LLMs, such as UX researchers and data scientists working for the development of LLMs, legal and risk teams for their assessment, product managers of consumer organization for their adoption, and end-users (cf. \autoref{fig:trustSummary}). We questioned them about their practices and challenges vis-a-vis the LLM, insisting on their relations with other practitioners in the supply chain, and discussing their opinions about LLMs and trustworthy AI. These interviews (over 3600 minutes of recording)  were reviewed and approved by the ethics committee of one of our institutions.
We then analyzed the interview transcripts deductively and inductively. We especially relied on aspects of trust discussed in prior literature above to inform our analysis and assess the novel or confirmatory character of our findings.
Note that the analysis of the transcripts is still in progress. In the future, we may identify more trust relations and trust factors, and nuances to those discussed in this paper, that would allow us to better disentangle the role trust plays in the production and adoption of LLMs.

\section{Preliminary Findings \& Discussion}


\subsection{Trustors \& Trustees: Trust Relationships Are Extremely Diverse Across the Supply Chain} 


\subsubsection{Many Trustors and Trustees Shape the Supply Chain} 
We find that, while the trustor is always a human entity, the trustee can be either a human entity or a technical artifact related to an LLM. 

\paragraph{Human trustors and trustees can be organizational or individual entities}
We find that the human entities in trust relationships can be of two different natures: individual or organizational entities.
For instance, the individual end-user of an LLM deployed by organization A might trust the LLM they use, or they might trust organization A to build a responsible LLM, or they could even indirectly trust the LLM because one of their friends might have used it extensively and argued for its usefulness profusely. 
Similarly, an organization B that decides to buy the LLM services of organization A typically develops a relation of trust with the LLM or organization A before their purchase decision.

\paragraph{Individual and organizational entities play diverse roles vis-a-vis the LLM, 
entailing different trust activities}
These entities all have one primary role vis-a-vis the LLM.\\ 
\textbf{\textit{Producer}}: An entity might participate in the development of any technical component that is used downstream to produce and maintain the final LLM (e.g., one might build the foundation model on which the LLM would be fine-tuned for a specific application). Producers often need to trust upstream components (e.g., the foundation model) to produce downstream ones (e.g., the fine-tuned model). Their primary vulnerability might be the reputational risk, in case a technical component or the LLM would be discovered to behave in a harmful manner or to rely on harmful premises --as this has happened in the past, e.g., testified by the OpenAI lawsuit \cite{Grynbaum_Mac_2023}. 
Note that it is often not the same organizational and individual entities working on the different 
components of the LLM. 
Policy documents \cite{madiega2021artificial,BSADoc} make the useful distinction between the ``developer'' and the ``deployer'' of the LLM, where the developer creates a fine-tuned LLM and the deployer makes this LLM ready for use, e.g., integrating it within a software stack, adding filters in the output of the LLM to circumvent potential offensiveness. We find even more fine-grained categories of roles at both the individual and organizational levels.\\ 
\textbf{\textit{Consumer}}: This can either be an organization deciding to use the services of an LLM and providing access to these services to its employees or external users, or it can be an independent user. The consumer employs the LLM to conduct a certain task for their personal or professional work, and typically requires some trust to adopt and continuously use the LLM with, e.g., the vulnerability that the LLM could provide them with offensive outputs. 
\\ 
\textbf{\textit{Indirect stakeholders}}: Other human entities, beyond those explicitly participating in the LLM supply chain,
might be impacted by the LLM or might simply have an opinion about the LLM and its impact on society. 
This could be considered a trust relation if the entity would consider themselves or others in society at risk due to the development or use of the LLM \cite{solaiman2023evaluating,kulynych2020pots}.

\paragraph{Trusting LLMs is more than trusting a stand-alone, shrinkwrapped, technical artifact}
As hinted above, an LLM is developed along multiple phases and relies on several technical components. We find that the trustee in trust relations might revolve around the final LLM or its separate components, either the components of the LLM system in use or the components that have served in building these in-use components. 
In use, an LLM-powered system is composed of a fine-tuned model for the application at hand, but also user interfaces for its end-users, workflows for post-filtering the outputs of the LLM (e.g., for toxicity), logging data and monitoring potential issues, infrastructures to support the computations and data needs, etc. Note that an LLM system might even be composed of several fine-tuned models chained together \cite{nushi2018towards}. 
Building each of these in-use components requires various build-up components, such as pre-training datasets, the foundation model on which the model was fine-tuned, a fine-tuning dataset, training scripts, data processing scripts, more infrastructures to handle data and computations, etc. 
Naturally, the granularity of these components is not set, and sub-components make up these in-use-components and build-up-components, e.g., a training dataset is made of data samples and annotations, a filtering workflow might be made of several filters for gender bias, race bias, toxicity, etc.
\textit{P33 ``
[Deployer organization], as a trusted platform, is easier to get started with and to deploy use cases on than building a new system from scratch, and having to work through your own security and firewalls and load balancers, and all the complexity behind standing up a new web application.''}

\subsubsection{Many Trust Relations Connect These Entities Across the LLM Supply Chain}
Based on all these entities in the LLM supply chain, we identify a diversity of trust relations that are marked by different trust activities. For each trust relation, we systematically disentangle the trustee(s), trustor(s) and trust activity(ies).

\paragraph{Downstream to Upstream, Entities Trust Different Technical Artifacts}
Entities relying on an LLM or one of its components, i.e., entities that are relatively downstream in the supply chain, often posed unexpected obstacles upstream due to potential issues (e.g., low performance and bias of technical artifacts), respectively for the adoption of the LLM, or for the adoption of its components and consequently for the development of the LLM itself.
\textit{P35 [consumer organization] ``There was a hesitancy to use it. There was a lack of trust. "I want to see it proven out before I bring it into my organization."  Now there's a stronger appetite to bring it in because they see the efficiency gains.''}
Hence, there, trust relations are not only between (1)~the individual end-user and the LLM, but also between the (2)~consumer organization and the LLM, as well as trust (3)~between the deployers and the fine-tuned model (and potential additional in-use components provided with the fine-tuned model), and trust (4)~between the developers and the build-up components they are given. In all cases, the positive expectation is for the technical components to work properly. The vulnerability consists in the repercussions that using a flawed AI system can have, e.g., while it might be reputational risk for an organization due to problematic outputs of the AI system or unethical ways to build it, it would be about the impact of problematic outputs (e.g., inaccurate or offensive) for the end users.
\textit{P70 LLM user ``
If it's wrong and I blame AI, that's gonna make me look unprofessional. So I try not to overuse it, it's a delicate balance. I'm probably going to get sued, or I'm gonna be the one in trouble. If I'm using AI as help, a lot of people don't trust that already, they want the Encyclopedia Britannica. I'd have to take responsibility for that. The big companies are probably gonna blame me.''}

\paragraph{Upstream Entities Trust Downstream Entities to Responsibly Use Technical Artifacts}
Our interviews also show that trust is bidirectional when talking about the usage of an LLM artifact. It is not only that the user of an artifact trusts the artifact or its developer (discussed above). It is also that the developer of the artifact trusts its consumer to use it appropriately (e.g., not outside the pre-defined scope of applications) to avoid any harm, at the risk of reputational issues. They also have to trust the consumers that they will not game the LLMs, e.g., via adversarial attacks to recover private information, and that they potentially put safeguards in place themselves. They might also trust the consumer organization to create trainings for its end-users to avoid misuse, e.g., not to over-rely on LLMs. In turn, the consumer organization trusts its individual end-users not to misuse the system and to properly handle problematic outputs of the system. For instance, one organization mentioned that when the end-users are internal, they trust their end-users to internally report on any toxic or offensive output and not to cause a public outcry and discuss the issue on social media. 
In summary, these upstream-to-downstream trust relations impact the responsible use of LLMs. 

\subsubsection{Discussion}
Our results confirm our intuition: the LLM supply chain depends on a diversity of trust relations. 
These trust relations involve many more trustors and trustees than current research on trust in AI envisions. The objects of these relations (i.e., the activity conducted by the trustee) also revealed to be more diverse than studied until now. 
This is aligned with prior research on trust in the context of automation (prior to AI) \cite{lee2004trust}, where one has investigated trust between individuals (``partner trust'', ``interpersonal trust''), between individuals and organizations (or even institutions), or between several organizations. 
Trust relations are often prerequisites to drive the production and adoption of LLMs. They are also vectors that support or hinder the establishment of trustworthy LLMs and their responsible development.
Closest to our findings is the study conducted by Browne et al. \cite{browne2022trust} on trust in clinical AI contexts. This study shows that clinicians (trustors) trust AI systems  (trustees) based on different activities all along the AI lifecycle (e.g., how AI training data were collected, which team developed the AI), and such trust is necessary for the clinicians to adopt and use the AI system. Yet, this study does not adopt a supply chain perspective. By accounting for the dynamics of the supply chain, our results further highlight the importance of inter-organizational trust, and identify trustors and trustees across the supply chain whose collaborative work impacts LLM production and adoption.

\subsection{Trustee-Related Factors: Trust is Fostered By Diverse Trust Cues About the Trustee} 

Our study revealed that the same types of factors as identified in prior works impact trust in the newly-surfaced relations. 
Therefore, through the lens of the ABI framework~\cite{mayer1995integrative}, for each entity (i.e., organizations, individuals, LLMs), we discuss the cues that might express their ability, benevolence, and integrity. 

\subsubsection{The ABI Properties of \textbf{Organizations} Target Trustors Differently}
Organizations, i.e., trustees, put forward cues related to their ability, integrity, and benevolence, 
particularly towards their customers, and put efforts into improving their actual trustworthiness in that regard. 
For instance, the hiring of researchers and publications of research papers, and cross-organization collaborations towards the development of LLMs \cite{Cholteeva_2023} provide a sense of ability, signed contracts and treaties \cite{Walker_2023}, responsible AI principles \cite{de2021companies}, and emerging governance structures (e.g., many participants mentioned that with the rise of LLMs, review boards and working groups are slowly established towards handling the risks that LLMs pose) instead illustrate benevolence and integrity.
Marketing campaigns and other release of communication pieces served to communicate these cues to certain trustors (particularly customers of the developer organizations).
On the other hand, we did not find any trustworthiness cues on the side of the organizations that adopt LLMs, neither towards their employees nor towards their end-users, nor any benevolence cues inter-organizational (e.g., around job displacement) and towards indirect stakeholders (e.g., artists whose content could be used as training data \cite{WashingtonPostEditorialBoard}). 

On the trustor's side, how these cues are perceived remains to be better understood. For instance, many of our participants, in their capacity as trustors, discussed the absence of benevolence from certain organizations, and instead emphasized the profit motive of these organizations. 
Additionally, some participants acknowledged benevolence but recognized the practical impossibility of exercising benevolence and developing meaningful governance processes in the race toward LLM development and adoption. For instance, while data stewards exist, whether they can comprehensively fulfill their duty and whether rigorous evaluations of their work are conducted and communicated (e.g., percentage of copyrighted data still included in training datasets) were different questions.

\subsubsection{Not All \textbf{Individual} Trustees Display All ABI Properties}
Many trustors discussed trust towards employees of the same or different organizations, referring to their expertise for building LLMs (ability), 
their integrity in building the LLMs, 
and general benevolence. Ability was referred to via the publication of academic papers in prestigious venues, corporate recognition, the belief that certain teams had the knowledge and could teach others about trustworthy LLMs, and the responsible AI-related communication efforts and actions organized by a few individuals. In terms of benevolence, participants discussed their own motivation and the dedication of their colleagues towards social good and developing ethical systems. 
Discussions about integrity only revolved around the end-users of LLMs. For instance, organizations employing these users or indirect stakeholders discussed users' integrity in terms of not querying LLMs with prompts to copy the styles of living artists. 
We did not identify any other integrity consideration for any other type of participant, nor questions of benevolence and ability from the consumers towards providers or indirect stakeholders.

\subsubsection{ABI Properties of \textbf{LLMs} Are Conflated With the Properties of Other Trustees} 
Research has long investigated the factors that make AI trustworthy and that communicate its trustworthiness \cite{liao2022designing}. In our interviews, the same factors and cues surfaced overall, particularly in terms of ability. 
\textit{P40 ``Everyone's gonna want something different. 
We're gonna get to a place where it is gonna come to trust. I trust that this is going to protect my information. I trust that this is going to be a good response. I trust this could be reliable. And if I can't trust it, I'm not going to use it, whether I'm the one that's supporting it, bringing it into the company or an end user. If I can't trust this response, I'm gonna call the help desk every single time.''}
We found that the benevolence of the LLM-powered system 
has been conflated with the benevolence of the producer organizations (particularly the deployer one) towards the consumer (similarly to \cite{liao2022designing,kim2023humans}), but also the benevolence of the consumer organization towards its individual users. 
In terms of integrity, LLM integrity revolves around ``the degree to which the operational or decision process of the model is appropriate to achieve the users’ goal'' \cite{liao2022designing}. The primary discussions we identified around this topic were about receiving indications about the documents used for the outputs of retrieval-augmented generation use-cases.
This was especially relevant to the consumer organizations, but we did not find such forms of AI integrity (nor others such as explainability) mentioned by any other participants.  

\subsubsection{Discussion}
Our results showed a plurality of factors that impact trust relations. These factors align with the ability, benevolence, integrity framework proposed by Mayer et al. \cite{mayer1995integrative} and applied to AI from a communication perspective by Liao and Sundar \cite{liao2022designing}. Naturally, the exact factors of trustworthiness and the cues that signaled these factors were however different across different trusted entities, and consequently different from those prior studies have investigated. 
This confirms the importance of disentangling the LLM supply chain, the entities that shape it, and the exact trust relations that traverse it, to more comprehensively identify trust factors. 
While these factors had not been explicitly discussed in the past in the context of AI, a few studies \cite{kim2023humans,ehsan2021expanding} have hinted at the impact of the organizational context on the appreciation of an AI system. Here, the ``organizational context'' is a trust relation where the trustee might not be the LLM but instead an employee or organization, and this relation impacts trust towards the LLM.


\subsection{Trustor-Related Factors: Trust, Appropriate Trust \& Reliance also Depend on the Trustor and Context}  

 \subsubsection{Trust Varies Across Trustors}
We notice that the extent of trust across trustors towards the same trustee varies. 
 We found that both contextual and personal factors mediate trust relations.
In terms of context, not only the stakes of the trust activity impact the relation (e.g., whether adopting the LLM might create financial risks or reputational damage to the consumer organization), but also the expectations the trustor has towards the trustee. 
For instance, we find that different consumer organizations have different expectations towards the ability of the LLM provided by the deployer organization depending on the environment in which individual users will use this system. Particularly, when individual users are employees of the consumer organization, we primarily find expectations in terms of accuracy, while in cases where the users are external to the organization, we also find expectations concerning the offensiveness of the outputs of the AI system.

In terms of personal factors, the natural propensity for trust in organizations and in technology, the ways individuals perceive AI systems, their advantages, but also their potential concerns, e.g., about dataset biases, also play a role.
Besides, the knowledge and awareness of the participants about LLMs, how they function, and what harm they can cause also impacted the types of trustworthiness cues they were looking for and their potential distrust towards LLMs.
\textit{P34 ``Continuing to educate everybody on what's the best use of AI and 
what is trustworthy AI, it will allow each team to make better calls of what to use.''} 

\subsubsection{The Supply Chain is Prone to (Un)Calibrated Trust} 
Reflecting on all the factors we surfaced, we found various avenues that lead to potentially non-calibrated trust. The ways trust relations affect each other is one of the key risk areas. On the side of the trustor, for instance, we found that both the temporality of the supply chain, the organizational versus interpersonal trust, and the impact of different trust activities can sometimes lead to blindly trusting an LLM despite its characteristics not meeting trustworthiness expectations. For instance, consumer organizations explained trusting deployer organizations for developing trustworthy traditional machine learning systems, and for this reason, they would also trust the deployers for developing trustworthy LLMs. 
\textit{P48 ``That's one of the things that establishes [developer-org] as an authority on it. We don't have this relationship with other vendors, so how do they expect adoption. So I think just engaging us in things like this really helps establish you as an authority. Maybe it's a bad thing, and it's kind of that segregation of duties: we've built this trust, and therefore we're going to trust you.
''}
Within chained trust relations, we also identified potentially wrong beliefs (e.g., about other's trust relations), that led to inappropriate trust. 

On the side of the trustee, we found that the trustee often needs the trust and reliance of the trustor for various reasons. Upstream, this is for job retention and climbing the corporate ladder, prompting the development of LLMs and their use and hence economic growth, and downstream for getting access to the technical artifacts. In this context, the trustee might have the incentive to develop unwarranted trust. This incentive sometimes plays a role in the beliefs of the trustor in the integrity of the trustee. It can also lead the trustee to display untruthful or misguiding trustworthiness cues \cite{corbett2023interrogating,birchall2011introduction,chromik2019dark}. 
\textit{P18 ``There's some sort of inherent trust that it is someone else's problem down the line, 
and that's why I go back to: the person writing the requirement needs to make sure that they're considering these things.''} 
Yet, the need for trust also becomes, in certain cases, a motivation for trustworthiness. Organizations recognizing that trustworthiness could become their competitive advantage compared to simply increasing the accuracy of an LLM might put more effort into developing such trustworthy LLMs. 
\textit{P15 ``If you throw solutions which lose customers, we lose everything. So we are very mindful.  We want to be responsible. We want trust. It's not just customers, even internal people.''}

Note that we also found situations where one entity relies on another one without trusting it. For instance, in the case of the user organization, they might want to buy the services of an LLM to remain competitive (vis-a-vis competitors who would also do so) while knowing that this LLM is not yet trustworthy. 

\subsubsection{Discussion}
These results show that trust is not only impacted by the characteristics of the trustee but also by those of the trustor and the context in which it is. While the characteristics of the trustor that impacts their trust relationship are overall not different from prior work (e.g., except their general propensity to trust organizations), the LLM supply chain brings new contextual factors of importance. 
To the best of our knowledge, no prior work has discussed these contextual factors. 
Finally, we found that these contextual factors and personal factors can lead to both calibrated and non-calibrated trust, and trust is not always present when reliance is. 
To the best of our knowledge, prior work \cite{chromik2019dark} had not identified the intentional (or not) ways to deceive trustors.
Furthermore, while prior works have made clear the distinction between trust as an attitude and the behaviors that can result \cite{lee2004trust}, accounting for the complexity of LLM supply chains has allowed us to identify and discuss organizational strings and powers that can impact trust and related behaviors.

\section{Implications \& Future Research}

Our preliminary results bear implications for practitioners, researchers, and policy-makers, and raise diverse research questions for each of these stakeholders. 

\paragraph{How Does Trust Fit in LLM Governance?}
Our findings showed the complexity of trust relations and hinted at the challenges toward establishing calibrated trust across the supply chain. Trust enables collaborations and collaborations are necessary for the 
development and adoption of LLMs. Furthermore, calibrated trust should enable appropriate collaborations, ideally leading to compliant and ethical LLM supply chains. 

This suggests the importance of bringing research on trust and AI collaborations to policy-makers and LLM governance researchers. Ideally, they should account for the variety of trustors and trustees across the supply chain, and for the challenges, needs, and driving forces each of them has. How can we account for trust relations while ensuring compliance of LLMs to their upcoming regulations? If it is possible to achieve a certain level of trust and appropriate trust within and across organizations, what is the most appropriate model of LLM supply chain governance? How do current notions of accountability \cite{raji2020closing}  and distributed responsibility \cite{orr2020attributions} fair with the complex AI supply chain we identified, and particularly with the various trust relations required in this supply chain?
Furthermore, future policies would need to acknowledge the various factors that might impact trust, and especially calibrated or unwarranted trust across the different entities in the LLM supply chain. Hence, how can we ensure that each trustee develops their ability, benevolence, and integrity appropriately, and that the trustworthiness cues they put forward are truthful? How do such trustworthiness cues fit with the potential trade secrets of each organization in the supply chain? From the trustor's side, do we need to foster more reflexivity among each entity of the supply chain, before trusting them to build compliant systems and interact with these systems appropriately? Or should we instead establish stricter requirements and assessment processes? Can internal assessments of LLMs be truthful if they rely on trust?
Prior research on trust and governance outside AI contexts could be a good start to answer these questions  \cite{gulati2008interorganizational}.

\paragraph{What Kind of Trust Is Desirable and Achievable?}
Our findings also suggest taking a step back and asking whether any type of trust is achievable and desirable anywhere in the supply chain, and what type of trust do we actually want to foster in each trust loci. 
On the one hand, subjectivity surfaced across the supply chain. 
We found that trustors have different expectations concerning the ability, integrity, and benevolence of an LLM and its supply chain. We also found that they expect different values to be accounted for in LLMs. 
Questions of subjective expectations for the outputs of LLMs and whether they can feasibly be met have already been raised \cite{kirk2023personalisation}. Research has further shown the impact of trustors sharing the trustee's values on trust \cite{mehrotra2021more}. Bringing these prior results to the LLM supply chain could hint at the potential impossibility of building trust across the entire supply chain.
This suggests that we should be cautious in defining the type of trust and trustworthiness cues needed across the supply chain. Instead of relying on each trustor's individual expectations, we might for instance need to foster discussions between supply chain entities to agree on a common set of expectations for the LLM supply chain, and to subsequently base trustworthiness cues on such commonly-accepted expectations. 
On the other hand, because the supply chain is so complex, it revealed impossible for each trustor to develop a well-informed trust in each relevant trustee: their visibility on upstream and downstream parts of the supply chain is limited \cite{widder2023dislocated}. As a result, we might want to distinguish different trust relations and different desirable processes to foster trust in these relations. For instance, we could investigate how to develop informed, reflexive, trust among entities that are close in the supply chain, while accepting ``blind'' trust among entities that are more distant as this is also necessary to the development of the LLM supply chain.

\paragraph{Studying New Trust Relationships in the LLM Supply Chain}
Finally, the novel character of our findings confirms the importance of adopting empirical, qualitative methodologies, grounded in the realities of the study participants to further investigate trust. With current assumptions in existing, quantitative, studies about trust in AI (trustors, trustees, activities, and impacting factors) \cite{mehrotra2023systematic}, we would not have been able to identify these new complexities.
Our study is however only the start of the exploration of trust within the LLM supply chain. 
We will need to investigate in more detail each of the trust relations and their impacting factors. For such research, we envision several practical and conceptual challenges. 
In terms of recruitment of participants, we do realize the difficulty of entering in contact with diverse stakeholders within organizations, and of discussing questions of trust that might be confidential. This might become especially challenging for quantitative studies that would require more participants. Yet, starting by conducting quantitative studies among populations of LLM users (as done currently) and AI developers before broadening to other entities, and accounting for the new factors, would already provide impactful insights. We also envision that the type of organization in which the participants work could impact their trust, and it will be important to understand the relevant dimensions of organizations (e.g., AI maturity \cite{rakova2021responsible}). Finally, researchers will need to account for the diverse designations that are used across organizations to refer to potentially similar or different roles vis-a-vis the LLM supply chain, and will also need to pay attention to terminological confusions in the way they address these different roles (varying professional cultures and jargons across organizational teams \cite{hall2005interprofessional}).

Another challenge will revolve around adapting the ways to measure trust, and the ways to present the impacting factors. To set up the studies about the trust of the user of an AI system, the reliance behavior on individual outputs of the AI system \cite{vereschak2021evaluate} is often used as a proxy. However, the new trust relations we identified do not necessarily involve reliance choices over these different outputs, but instead might revolve around the decision to adopt or to provide access to a technical artifact, or to delegate such decision to another, potentially more knowledgeable, individual in the same organization. Therefore, whether and how these behaviors could be used as new proxy measures remains to be investigated. 
Additionally, careful thought should also be put into developing ways to convey potential relevant factors to the trustees in the studies. Particularly, this will require understanding how to present the values of the organizations within the supply chain (e.g., presenting them vaguely, in an explicit list, in a ranked list, with or without information about the purpose of the LLM), as our study showed that the participants not only weigh various advantages and disadvantages of LLMs differently, but also prioritize values differently (similarly to \cite{jakesch2022different}), and that these prioritizations might depend on the purpose of the LLM and the stakes they have in it. 



\bibliographystyle{ACM-Reference-Format}
\bibliography{main}

\end{document}